\documentclass[11pt,tightenlines,eqsecnum,floats,aps,amsmath,amssymb,nofootinbib,prd,shownopacs,notitlepage,longbibliography]{revtex4-2}
\usepackage{graphicx}
\usepackage[utf8]{inputenc}
\usepackage{amsmath,amsfonts,amssymb}
\usepackage{color}
\usepackage[hidelinks]{hyperref}
\hypersetup{
  colorlinks   = true,
  urlcolor     = blue, 
  linkcolor    = blue,
  citecolor   = blue 
}
\begin{document}

\title{Constraining dark energy models using Jackknife and Bootstrap resampling}
\author{Roshna K}
\email{roshnak.217ph005@nitk.edu.in }
\author{Nikhil Fernandes}
\author{P Praveen}
\author{V. Sreenath}
\email{sreenath@nitk.edu.in}
\affiliation{Department of Physics, National Institute of Technology Karnataka, Surathkal, Mangaluru 575025, India.}
%%%%%%%%%%%%%%%%%%%%%%%%%%%%%%
\begin{abstract}
Analyses of type Ia supernovae have helped us shed light on the existence and nature of dark energy. 
Most of these analyses have relied on Bayesian techniques. In this work, we employ resampling techniques, namely Jackknife and Bootstrap, together with generalised least squares, to analyse supernova data. We first calibrate these techniques using near-ideal mock data and versions of the PantheonPlus data, and compare their performance with Bayesian methods.
We find that with near-ideal mock data, Jackknife can yield better constraints. We then apply these methods to constrain  parameters of flat $\Lambda$CDM, $\Lambda$CDM, flat $w$CDM, $w$CDM, and flat $w_0\,w_a$CDM models from the PantheonPlus and SH0ES (PPS) data. We observe that constraints obtained with different techniques for three- and four-parameter models are largely consistent. We also find that Hubble tension is less significant with constraints from Jackknife. For instance, we find that Jackknife estimates $h\,=\,0.743\,\pm\,0.029$ from PPS data, making the Planck value well within $3\sigma$. Moreover, we estimate $h\,=\,0.678\,\pm\,0.052$ from PantheonPlus data when we only consider sources with $z > 0.01$, 
in which case there is no tension with the Planck estimate. 
These results highlight the importance of using multiple techniques while analysing data and warrant further investigation.
\end{abstract}

\maketitle
\section{Introduction}\label{sec:1}
Observations of distant Type Ia supernovae (SNe Ia) have provided the first evidence for the late-time accelerated expansion of our Universe \cite{SupernovaCosmologyProject:1998vns, SupernovaSearchTeam:1998fmf}. 
This accelerated expansion is attributed to dark energy, which currently dominates the universe's energy density. The exact nature of dark energy remains a mystery. 
The simplest explanation for dark energy is that it is a cosmological constant $\Lambda$, described as a constant energy density with negative pressure, corresponding to an equation-of-state parameter $w = -1$ \cite{Peebles:2002gy}. Together with the idea of cold dark matter, it leads to the standard model of cosmology, known as the $\Lambda$CDM model. 
\par  
Despite the success of the $\Lambda$CDM model in explaining a wide range of observations, recent measurements of Baryon Acoustic Oscillations (BAO) data provided by the Dark Energy Spectroscopic Instrument (DESI) Collaboration, in combination with the Cosmic Microwave Background (CMB), have revealed tensions with that of cosmological constant-like dark energy \cite{DESI:2025zgx}. These discrepancies suggest the presence of a dynamical dark energy, wherein the equation of state parameter varies with the redshift. Various parametrisations of the dynamical dark energy equation of state parameter $w(z)$ have been proposed in the literature (see, for instance, \cite{DiValentino:2021izs}). One of the simplest parametrisations of the dynamical dark energy equation of state parameter is a constant $w(z)$, but not necessarily equal to $-1$ (see, for instance, \cite{An:2016keq, DiValentino:2017gzb, Marcondes:2017vjw, Escamilla:2023oce}). This model of dark energy is referred to as the $w$CDM model. A more general parametrisation of the equation of state parameter $w(z)$ is the Chevallier-Polarski-Linder (CPL) parametrisation, in which $w(z)$ slowly varies linearly with redshift \cite{Chevallier:2000qy, Linder:2002et}. This simple and well-known two-parameter model of dark energy is called the $w_0 w_a$CDM model. In this model, the equation of state parameter $w$ evolves with redshift as $ w(z) = w_0 + w_a \frac{z}{1 + z}$, where $w_0$ denotes the present value of the equation of state parameter and $w_a$ describes its evolution with the redshift. 
\par 
In this work, we investigate these models of dark energy and constrain their model parameters using the state-of-the-art PantheonPlus and SH0ES supernovae dataset \cite{Scolnic:2021amr, Brout:2022vxf}. While several studies \cite{Reboucas:2024smm, Cheng:2025lod, Sharma:2025rmj, Barua:2025ypw} have already provided Bayesian constraints for these models, our goal is to further examine the consistency and reliability of these results by applying different statistical methods. In this work, we utilize non-parametric resampling techniques \cite{Efron_bradley, efron_tibshirani, Shao1996TheJA} such as the Jackknife and Bootstrap methods together with the Generalised Least Squares (GLS) method. These resampling techniques involve repeatedly drawing samples from the original dataset and estimating the model parameters. The main advantage of these techniques is that they do not rely on any assumptions about the underlying distribution of the data.
In addition to the above frequentist methods, we also
analyse the dark energy models using Bayesian approaches.
In the Bayesian framework, we perform analysis using two different sampling algorithms: the Markov chain Monte Carlo (MCMC) \cite{Hastings:1970aa}, and the advanced nested sampling algorithm \cite{Skilling:2006gxv}.
\par 
The paper is organized as follows. In Section \ref{sec:background}, we present the theoretical background and describe the various models of dark energy considered in this work. In Section \ref{sec:dataset}, we describe the PantheonPlus and SH0ES dataset used in our analysis. This is followed by Section \ref{sec:techniques}, where we discuss the statistical methods, including frequentist analyses such as GLS and resampling techniques, namely Jackknife and Bootstrap, as well as the Bayesian approach, including MCMC and nested sampling, employed in this study. 
In Section \ref{sec:calibration}, we calibrate these techniques using a mock dataset as well as with different versions of PantheonPlus data.
We present the constraints obtained for each cosmological model using the different methods with the PantheonPlus and SH$0$ES dataset in Section \ref{sec:constraints}. In Section \ref{sec:discussion}, we conclude the paper with a summary and discussion of our results.
%%%%%%%%%%%%
\section{Theoretical background and models of dark energy}\label{sec:background}
We consider the homogeneous and isotropic Friedmann–Lema\^itre–Robertson–Walker (FLRW) geometry with the metric, 
\begin{eqnarray}
ds^2 = c^2\,dt^2 - a^2(t) \,\bigg( \frac{dr^2}{1 - \kappa\,r^2} + r^2\,(d\theta^2 + \sin^2 \theta\,d\phi^2)\bigg),
\end{eqnarray}
where $a(t)$ is the scale factor that captures the expansion of the Universe, $\kappa$ is the spatial curvature that takes values of $-1, 0, 1$ for an open, flat, and closed Universe. The expansion rate of the Universe is described by the Hubble parameter $H(z)$ with its present value given by the Hubble constant $H_0$. Its evolution with redshift is given by,  
\begin{equation}
    H(z) = H_{0} \sqrt{ \Omega_r (1+z)^4 + \Omega_m (1+z)^3 +  \Omega_k (1+z)^2 + \Omega_{\rm DE} f(z)},
\end{equation}
where $\Omega_r$, $\Omega_m$, $\Omega_{\rm DE}$, and $\Omega_k$ represent the density parameters for radiation, matter, dark energy, and curvature, respectively, evaluated at the present time. The function $f(z)$, which encodes the dynamics of dark energy, is related to the equation of state parameter $w(z)$ as (see, for instance, \cite{Padmanabhan_2002}),
\begin{equation}
    f(z) = \exp \left[ 3 \int_0^z \frac{1 + w(z')}{1 + z'} \, dz' \right].
\end{equation}
The density parameters satisfy the condition $\Omega_r + \Omega_m + \Omega_k + \Omega_{\rm DE} = 1$. In the late universe, contributions from the radiation are negligible. Hence, we will approximate $\Omega_k = 1 - \Omega_m - \Omega_{\rm DE}$. For our analysis, we consider both spatially flat and non-flat cosmological models with different parametrisations of dark energy. We describe these models below. 
\subsection{$\Lambda$CDM model}
In the standard $\Lambda$CDM model, dark energy is described by a cosmological constant with an equation of state $w = -1$. The Hubble parameter for this model is given by,
    \begin{equation}\label{eqn:lcdm}
        H(z) = H_{0} \sqrt{ \Omega_m (1+z)^3 + (1 - \Omega_m -\Omega_{\rm DE})(1+z)^2 + \Omega_{\rm DE} }\,, 
    \end{equation}
where spatial curvature $\Omega_k$ is written as $1 - \Omega_m -\Omega_{\rm DE}$. The free parameters in this model are $\Omega_m$, $\Omega_{\rm DE}$ and $H_0$. 
\subsection{Flat $\Lambda$CDM model}
If we assume a spatially flat Universe with $\Omega_k = 0$, the Hubble parameter given in Eqn. \ref{eqn:lcdm} reduces to
    \begin{equation}
        H(z) = H_{0} \sqrt{ \Omega_m (1+z)^3 + (1 - \Omega_m) }\, ,
    \end{equation}
where $\Omega_{\rm DE}$ is written as $1 - \Omega_m$. The unknown parameters in this model are $\Omega_m$ and $H_0$. 
\subsection{$w$CDM model}
    In the $w$CDM model, the equation of state parameter $w(z)$ is constant and does not evolve with redshift. This is different from the $\Lambda$CDM model, where $w$ is fixed to be $-1$. In this model, although the equation-of-state parameter is independent of redshift, it can take any value, $w(z) = w_0$. The evolution of dark energy is given by the form,
    \begin{equation}
        f(z) = (1 + z)^{3(1 + w_{0})},
    \end{equation}
    and hence the Hubble parameter corresponding to this model is given by,
    \begin{equation}\label{eqn:wcdm}
        H(z) = H_{0} \sqrt{ \Omega_m (1+z)^3 + (1 - \Omega_m -\Omega_{\rm DE}) (1+z)^2 + \Omega_{\rm DE} (1+z)^{3(1+w_{0})} }.
    \end{equation}
In this model, there are four free parameters, namely $\Omega_m$, $\Omega_{\rm DE}$, $w_0$ and $H_0$. 
    \subsection{Flat $w$CDM model}
    For a spatially flat universe with $\Omega_k = 0$, the Hubble parameter given in Eqn. \ref{eqn:wcdm} becomes, 
     \begin{equation}
        H(z) = H_{0} \sqrt{ \Omega_m (1+z)^3 + (1 - \Omega_m)(1+z)^{3(1+w_{0})} }.
    \end{equation}
The free parameters in this model are $\Omega_m$, $w_0$, and $H_0$. 
   \subsection{Flat $w_0\,w_a$CDM model}
    In this model, the equation of state of dark energy evolves with redshift as,
    \begin{equation}
        w(z) = w_0 + w_a \frac{z}{1 + z}, 
    \end{equation}
    where $w_0$ denotes the present-day value of the equation of state and $w_a$ describes its evolution with the redshift. This parametrisation is also known as the CPL parametrisation \cite{Chevallier:2000qy, Linder:2002et}. The dark energy evolution function corresponding to this $w(z)$ is 
    \begin{equation}
        f(z) = (1 + z)^{3(1 + w_0 + w_a)} \exp\left[-\frac{3 w_a z}{1 + z}\right],
    \end{equation}
    and the Hubble parameter is given by
    \begin{equation}
        H(z) = H_{0} \sqrt{ \Omega_m (1+z)^3 + (1 - \Omega_m) (1 + z)^{3(1 + w_0 + w_a)} \exp\left[-\frac{3 w_a z}{1 + z}\right] }.
    \end{equation}
In this model, $\Omega_m$, $w_0$, $w_a$ and $H_0$ are the free parameters. 
%%%%%%%%%%%%%%%%%%%%
\section{Observational dataset}\label{sec:dataset}
In this work, we use the PantheonPlus compilation of Type Ia supernovae, an updated successor to the original Pantheon dataset, which is a comprehensive collection of 18 surveys of supernovae\footnote{\href{https://github.com/PantheonPlusSH0ES/DataRelease}{https://github.com/PantheonPlusSH0ES/DataRelease}}.
This dataset includes a sample of 1701 supernovae over the redshift range $0.001 < z < 2.261$, including those in galaxies with measured SH0ES Cepheid distances  \cite{Scolnic:2021amr, Brout:2022vxf}. The dataset also provides the associated measurement errors, represented by a full covariance matrix that combines both statistical and systematic uncertainties. 
This includes all covariance between SNe Ia and covariance between Cepheid hosts due to systematic uncertainties.
We refer to this PantheonPlus and SH0ES dataset as PPS. 

\par
For each supernova, the theoretical distance modulus $\mu(z)$, which quantifies the difference between an object's apparent magnitude (observed flux $m$) and absolute magnitude (intrinsic brightness $M$), is defined as \cite{Sahni:2004ai, Padmanabhan:2005ur,Copeland:2006wr},
\begin{eqnarray}
\mu_{\rm theory}(z) &=& m - M = 5\,{\rm log_{10}}\bigg(\frac{D_L(z)}{\rm Mpc}\bigg) + 25 ,
\end{eqnarray}
where $D_L(z)$ is the luminosity distance to a supernova with redshift $z$. To treat both spatially flat and non-flat cosmological models, the luminosity distance is expressed in its most general form \cite{Carroll:2000fy},
\begin{equation}
D_L(z) = \frac{c(1+z)}{H_0} \times 
\begin{cases}
\frac{1}{\sqrt{|\Omega_k|}}\sinh \left( \sqrt{|\Omega_k|} \int_0^{z} \frac{dz'}{E(z')} \right), & \text{if } \Omega_k > 0, k = -1 \text{ (open)} \\[1.2em]
\int_0^{z} \frac{dz'}{E(z')}, & \text{if } \Omega_k = 0, k = 0 \text{ (flat)} \\[1.2em]
\frac{1}{\sqrt{|\Omega_k|}}\sin \left(\sqrt{|\Omega_k|} \int_0^{z} \frac{dz'}{E(z')} \right) & \text{if } \Omega_k < 0, k = +1 \text{ (closed),}
\end{cases}
\end{equation}
where $E(z) = \frac{H(z)}{H_0}$ is the normalised Hubble parameter. 
To account for the full redshift range of PPS dataset, the theoretical distance modulus is evaluated using $z_{\rm HD}$, the redshift corrected for CMB dipole motion and peculiar velocity \cite{Dainotti:2024gca}.
For a given set of model parameters $\theta$, the theoretical prediction of the apparent magnitude is, 
\begin{eqnarray}\label{eqn:mtheory}
m_{\rm theory}(z, \theta) = 5\,{\rm log_{10}}\left[
\frac{3000\times(1 + z)}{h} \times
\begin{cases}
\dfrac{1}{\sqrt{|\Omega_k|}}\sinh \!\left( \sqrt{|\Omega_k|} \displaystyle\int_0^{z} \frac{dz'}{E(z')} \right) \\[1.2em]
\displaystyle\int_0^{z} \frac{dz'}{E(z')} \\[1.2em]
\dfrac{1}{\sqrt{|\Omega_k|}}\sin \left( \sqrt{|\Omega_k|} \displaystyle\int_0^{z} \frac{dz'}{E(z')} \right) \\[1.2em]
\end{cases}
\right]\,+\,25\,+\,M,\nonumber
\\
 && 
\end{eqnarray}
where $h = H_0\,/\,(100 \rm\, km\,s^{-1}\, Mpc^{-1})$ is the dimensionless Hubble constant. Note that, on the right-hand side of the Eqn. \ref{eqn:mtheory}, the parameters $\theta$ enters implicitly via the form of $E(z)$ as well as through $\Omega_k$ and $h$. 
For SNe Ia present in the \texttt{$\rm SH0ES$} Cepheid galaxies, instead of the theoretical apparent magnitude given in Eqn. \ref{eqn:mtheory}, we directly use the Cepheid distance modulus $\mu^{\rm Cepheid}$ given as \texttt{CEPH\_DIST} in PPS dataset. Hence, the theoretical prediction of the apparent magnitude is \cite{Brout:2022vxf}, 
\begin{eqnarray}\label{eqn:mtheory_ceph}
    m_{{\rm theory}}(z_i, \theta) = 
    \begin{cases}
M + \mu_i^{\rm{Cepheid}}, & i \in \text{Cepheid hosts}, \\[6pt]
m_{\mathrm{theory}}(z_i,\theta), & \text{otherwise}.
\end{cases}
\end{eqnarray}
The SH0ES Cepheid distances are included with the hope that it will help to break the degeneracy between the absolute magnitude $M$ and the Hubble constant $H_0$, which cannot be resolved with supernovae data alone. In our analysis, along with the model-dependent parameters discussed in the section \ref{sec:background}, we also treat the absolute magnitude of Type Ia supernovae, $M$, as an additional free parameter. 

%%%%%%%%%%%%%%%%%%%%%%%%%%%%
\section{Statistical techniques}\label{sec:techniques}
In this section, we discuss the statistical methods used to estimate model parameters along with the nuisance parameter $M$. We perform five different analyses, including methods from both the frequentist and Bayesian approaches.  Within the frequentist approach, we apply the GLS\footnote{Since we are applying GLS to a non-linear function Eqn. \ref{eqn:mtheory_ceph}, we may also call it a non-linear generalised least square. } method and two resampling techniques, namely the Jackknife and Bootstrap methods. To compare and complement these results, we also conduct two Bayesian analyses using the MCMC and nested sampling algorithms.

%%%%%%%%%%%%%%%%%%%%%%%%%%
\subsection{Generalised least squares }\label{subsec:chi_square}
This is a standard frequentist statistical method used to get the best-fit parameters of any given model. The fit between theory and data is assessed using the $\chi^2$ function (see, for instance, \cite{Barlow:0471922951}), which measures the difference between the theoretical predictions and the observational data. The parameters that best fit the data are identified by minimising the $\chi^2$ values defined as,  
\begin{eqnarray}\label{eqn:chisquare}
    \chi^2(\theta)&=&\Delta\,m^TC^{-1}\Delta\,m,
\end{eqnarray}
where $\Delta\,m = m_{\rm theory}(z, \theta) - m_{\rm observed}(z)$ is the residual vector, and $C$ is the  covariance matrix of the data.
For a given model of dark energy with a set of parameters, we can compute the theoretical apparent magnitude $m_{\rm theory}$ using Eqn. \ref{eqn:mtheory_ceph}. For each parameter $\theta_\alpha$, the best fit value of the parameter $\hat{\theta}_\alpha$ \cite{Padmanabhan:2002vv} are obtained by minimising the Eqn. \ref{eqn:chisquare}, such that 
\begin{equation}
    \left. \frac{\partial \chi^2}{\partial \theta_\alpha} \right|_{\theta_\alpha = \hat{\theta}_\alpha} = 0.
\end{equation}
To perform this minimisation, we employ the \texttt{Py-BOBYQA} algorithm \cite{Cartis:2018xum}, a derivative-free optimiser for bound-constrained minimisation problems. It is a Python implementation of the Bound Optimization BY Quadratic Approximation (\texttt{BOBYQA}) solver. The optimization is carried out within the parameter bounds listed in the table \ref{T:prior}.
To estimate the uncertainties associated with the best-fit parameters, we find the curvature matrix $A$, which is the second derivative of the $\chi^2$ function with respect to model parameters \cite{Padmanabhan:2002vv}, 
\begin{eqnarray}
    A_{\alpha \beta} = \frac{1}{2}\bigg(\frac{\partial^2 \chi^2}{\partial \theta_\alpha\partial \theta_\beta}\bigg).
\end{eqnarray}
 The elements of the curvature matrix $A_{\alpha \beta}$ are evaluated numerically using the central difference formula around the best-fitting parameter. The covariance matrix, which represents the pairwise covariances between each parameter, is then obtained as the inverse of the curvature matrix $C\, =\, A^{-1}$. The square root of the diagonal elements of $C$ gives the standard deviation associated with each parameter $(\sigma_\alpha^{\rm_{GLS}})$. In this method, best-fit parameter estimates are obtained without assigning any underlying probability distributions. However, associated variances are estimated by assuming an approximately Gaussian distribution around the minimum. 
 \par
To plot the confidence regions for any subset of two parameters $(\theta_\alpha, \theta_\beta)$, we evaluate the quantity 
\begin{eqnarray}
    \Delta\chi^2_{p}(\theta_\alpha, \theta_\beta)  = \sum_{\alpha, \beta = 1}^{2}(\theta_\alpha - \hat{\theta}_\alpha)A_{\alpha \beta}(\theta_\beta - \hat{\theta}_\beta), 
\end{eqnarray}
where $\alpha$ and $\beta$ vary over the two parameters.
The quantity $\Delta\chi_{p}^2(\theta_\alpha, \theta_\beta)$ is computed on a two-dimensional grid mapping the plotting range of parameters, and confidence contours are drawn at fixed values of $\Delta\chi_{p}^2$. 
Since $\Delta\chi_{p}^2$ follows a chi-square distribution, for two parameters, the $68\%$ and $95\%$ confidence regions correspond to a $\Delta\chi_{p}^2 = 2.30$ and $5.99$, respectively.
The one dimensional probability distribution for each parameter $\theta_\alpha$ is approximated as a Gaussian and is given as 
\begin{eqnarray}
    P(\theta_\alpha) \propto \exp\left[-\frac{1}{2} A_{\alpha \alpha}\left(\theta_\alpha - \hat{\theta}_\alpha\right)^2\right].
\end{eqnarray}

%%%%%%%%%%%%%%%%%%%%%%%
\begin{table}
    \centering
    \begin{tabular}{|c|c|}
    \hline 
    Parameter & Prior \\
    \hline 
       $\Omega_m$  &  [0.01, 0.9]\\
       $\Omega_{\rm DE}$  & [-2, 2]\\
       $h$ & [0.55, 0.91]\\
       $w_{0}$ & [-4, 4] \\
       $w_a$ & [-4, 4]\\
       $M$ & [-20, -18]\\
       \hline 
    \end{tabular}
    \caption{The lower and upper bounds of model parameters. In the frequentist approach, we limit the search for the best-fitting parameter to this range. In the Bayesian approach, we assume a uniform prior over this range. 
    \label{T:prior}}
\end{table}
%%%%%%%%%%%%%%%%%%%%%%%%%%%%%%
\subsection{Non-parameteric resampling techniques}
%%%%%%%%%%%%%
Non-parameteric resampling techniques are frequentist statistical methods in which we compute the estimates of interest by creating multiple copies of the original data, either by leaving out data points or by sampling with replacement. These techniques help us to explore how the results change when we slightly alter the original dataset. These resampling techniques have the added advantage of not requiring assumptions about the data's underlying distribution. 
In this work, we use two resampling algorithms, the Jackknife and the Bootstrap. Using these resampling techniques, we can compute parameter estimates and assess their uncertainty and bias.

%%%%%%%%%%%%%%%%%%%%
\subsubsection{Jackknife resampling technique}
%%%%%%%%%%%%
The Jackknife method is a simple resampling technique in which we remove data point(s) from the original dataset without replacement \cite{10.1214/aoms/1177729989, 6c956df0-ca97-3419-9961-dcc097853946}. This method assesses the stability of an estimator and quantifies its bias and variance.  In the standard jackknife delete-$1$ method, $N$ number of Jackknife samples are created, where $N$ is the total number of supernovae in the PPS dataset ($N=1701$). Jackknife resamples are created by systematically removing one observation at a time from the original PPS dataset without replacement. Hence, each of these Jackknife samples contains $N-1$ data points. For each of the Jackknife samples, we compute our estimate of the vector of best fit parameters
$\widetilde{\theta_s}$ using GLS method described in subsection \ref{subsec:chi_square}, and then combine all such estimates to obtain the Jackknife estimate. The vector of Jackknife estimate of parameters ${\hat\theta_{\rm jack}}$ \cite{Hogg:2010yz}, is given by, 
\begin{eqnarray}\label{eqn:jack_mean}
    {\hat\theta_{\rm jack}} = \frac{1}{N} \sum_{s = 1}^N \widetilde{\theta_{s}},
\end{eqnarray}
where $\widetilde{\theta_s}$ is the best fit parameter estimate obtained with each Jackknife subsample. A key aspect of the Jackknife method is that it provides an estimate of the bias made using full dataset.
The Jackknife estimate of the bias \cite{mcintosh2016jackknifeestimationmethod} is computed as 
\begin{eqnarray}\label{eqn:bias}
\widehat{\rm bias_{_{jack}}} &=& (N-1)\, (\hat\theta_{\rm jack} - \hat{\theta}), \end{eqnarray}
where $\hat{\theta}$ is the mean estimate obtained from the original PPS full dataset.  
The bias is scaled by the factor $N-1$ because each Jackknife sample contains $N-1$ data points.
If we have the estimate of bias, the Jackknife corrected estimate of the parameter of interest is given by, 
\begin{eqnarray}\label{eqn:jack_corrected}
    \hat\theta_{\rm jack}^{\rm{corrected}} = \hat{\theta} - \widehat{\rm bias_{_{jack}}}. 
\end{eqnarray}
 The Jackknife covariance matrix is given by \cite{10.1214/aos/1176345462}, 
\begin{eqnarray}\label{eqn:cov_jack}
    C_{\rm jack} = \frac{N - 1}{ N} \sum_{s = 1}^{N}(\widetilde{\theta_{s}} - \hat\theta_{\rm jack})(\widetilde{\theta_{s}} - \hat\theta_{\rm jack})^T. 
\end{eqnarray}
The corresponding Jackknife curvature matrix $A_{\rm jack}$ is the inverse of the covariance matrix $C_{\rm jack}$. The Jackknife estimate of the standard deviation of each parameter $\sigma_\alpha^{\rm{jack}}$ is given by the square root of the corresponding diagonal element of covariance matrix.
To plot the confidence contours for the subset of two parameters $(\theta_\alpha, \theta_\beta)$ in the Jackknife method, we follow a method similar to that adopted in GLS analysis. In this method, we use the covariance matrix given in Eqn. \ref{eqn:cov_jack} and confidence levels are determined using the $F$ distribution  rather than the $\chi^2$ distribution. 
For a set of two parameters, the contour levels corresponding to $68\%$ and $95\%$ confidence regions are obtained from the $F$ distribution with the numerator degree of freedom $\nu_{1} = 2$ and the denominator degree of freedom $\nu_2 = N - 2$, where $N$ is the number of Jackknife samples. 
The one dimensional probability distribution of a parameter $\theta_\alpha$ are plotted as a normal distribution with 
mean $\hat\theta_{{\rm jack},\,\alpha}^{\rm{corrected}}$ and standard deviation $\sigma_\alpha^{\rm{jack}}$.
\par 
We also implement a more general version of the Jackknife, namely the delete-$d$ method, in which $d$ data points are removed from the dataset. Each sample contains $(N-d)$ data points. We work with $d\,=\, \sqrt{N}$ and $d\,=\,N/10$. 
In practice, we generate $M$ number of such realisations ($M = 500$). %
The corresponding bias estimate is calculated by replacing the normalisation factor $N-1$ in Eqn. \ref{eqn:bias} to $(N-d)/d$. The corresponding covariance matrix of the parameters is estimated as
\begin{eqnarray}
C_{\rm jack} = \frac{N-d}{d \times M} \sum_{s=1}^{M}
(\widetilde{\theta_{s}} - \hat\theta_{\rm jack})(\widetilde{\theta_{s}} - \hat\theta_{\rm jack})^T. 
\end{eqnarray}
%%%%%%%%%%%%%%%%%%%%%%%%%%%%%%%%%%
\subsubsection{Bootstrap resampling technique}
 %%%%%%%%%
The Bootstrap technique \cite{Shao1996TheJA, 10.1214/aos/1176344552} involves generating $B$ number of new Bootstrap samples, each of size $N$, which is 1701, by sampling with replacement from the original PPS dataset. This means that in each Bootstrap sample, some data points may repeat, while others may be left out. For each of the $B$ Bootstrap samples, we compute the best fit parameter estimates using the GLS method described in subsection \ref{subsec:chi_square}. 
The Bootstrap mean estimate of the set of parameters {$\hat\theta_{\rm boot}$} \cite{Hogg:2010yz}  is given by, 
\begin{eqnarray}
\hat{\theta}_{\rm boot} = \frac{1}{B} \sum_{s=1}^{B}\widetilde{\theta}_s, 
\end{eqnarray}
where $\widetilde{\theta}_s$ is the vector of best fit values obtained for each of the Bootstrap samples. The Bootstrap also provides a way to estimate the bias of the estimator and is given as, 
\begin{eqnarray}
    \widehat{\rm bias_{boot}} = \hat{\theta}{_\mathrm{boot}} - \hat{\theta}. 
\end{eqnarray}
The Bootstrap corrected estimate is given by
\begin{eqnarray}\label{eqn:bias_corrected_boot}
     \hat\theta_{\rm boot}^{\rm{corrected}} = \hat{\theta} - \widehat{\rm bias_{_{boot}}}.
\end{eqnarray}
The Bootstrap estimate of the standard deviation is,  
\begin{eqnarray}\label{eqn:boot_variance}
\sigma_\alpha^{\rm boot} = \sqrt{\frac{1}{B} \sum_{s=1}^{B} (\widetilde{\theta}_{s,\,\alpha} - \hat{\theta}_\alpha)^2},
\end{eqnarray}    
where the subscript $\alpha$ refers to a parameter of interest and $\hat{\theta}_\alpha$ is the best fit value obtained from the original full PPS dataset.
In our analysis, we generated $B = 1000$ Bootstrap samples. In this method, we directly use the distribution of parameters from the $B $ Bootstrap samples, after correcting for bias, to plot the confidence contours and do not use the covariance matrix as done in GLS or in the Jackknife method.  For any subset of two parameters $\theta_\alpha$ and $\theta_\beta$, the joint probability contours are estimated using a kernel density estimation (KDE) approach. We use the KDE function available in the \texttt{seaborn} module. To plot the one-dimensional distribution of each parameter, we estimate the probability density using a Gaussian kernel density estimator, which is implemented using \texttt{gaussian\_kde} available in \texttt{SciPy}, and we work with the default bandwidth. 
%%%%%%%%%%%%%%%%%%%%%%%%%%%%%%%%%
\subsection{Bayesian analysis: MCMC and Nested sampling}
In the previous subsections, we discussed the frequentist approach, including the GLS and the two resampling techniques.
In this section, we describe parameter estimation using Bayesian methods. 
Bayes' theorem provides a systematic way to estimate the model parameters and their associated uncertainties and is stated as (see, for instance, \cite{sivia_book}), 
\begin{eqnarray}
    P(\theta|D) = \frac{P(D|\theta)\,P(\theta)}{P(D)},
\end{eqnarray}
where $P(\theta|D)$ is the posterior probability of the model parameters given the data $D$, $P(D|\theta)$ is the likelihood which quantifies how probable it is to observe the data given the underlying model parameters, $P(\theta)$ is the prior information about the parameters, and $P(D)$ is the normalization constant or evidence. 
If we assume a Gaussian-distributed dataset, the likelihood is expressed in terms of the 
$\chi^2$ as, 
\begin{equation}
    \mathcal{L}(\theta) \propto \exp\left(-\chi^2(\theta)/2\right).
\end{equation}
Thus, in this context, minimizing $\chi^2(\theta)$ is equivalent to maximizing the likelihood function. To perform the Bayesian analysis, we use two different sampling techniques, namely MCMC and the nested sampling \cite{Skilling:2006gxv}. In the MCMC sampling method, we sample the parameter space and obtain the posterior probability distribution of the model parameters. The sampler generates multiple sequences of points in the parameter space, known as Markov chains. We use the \texttt{emcee} Python package \cite{emcee}, which is an ensemble sampler to explore the parameter space. In this work, we assume flat priors for all model parameters as listed in table \ref{T:prior}. For our analysis, we use 16 chains, each with $40000$ steps. We use the \texttt{GetDist} package \cite{Lewis:2019xzd} to analyze the chains after removing the first 30\% as initial burn-in. The marginalized mean values and the associated $1\sigma$ uncertainties are computed with the remaining samples. The convergence of the chains is verified using the \texttt{Gelman-Rubin} statistic \cite{Gelman:1992zz}. We also perform the analysis using the advanced nested sampling algorithm, \texttt{PolyChord} \cite{Handley:2015fda, 2015MNRAS.453.4384H}. \texttt{PolyChord}  is designed for exploring higher-dimensional parameter spaces and can efficiently handle multimodal probability distributions. This algorithm works by drawing a set of $n$ live points uniformly from the prior. At each iteration, the live point with the lowest likelihood is removed and replaced by a new live point with a higher likelihood. We implement \texttt{PolyChord} through its Python interface, {\texttt{pypolychord} \footnote{\href{https://github.com/PolyChord/PolyChordLite}{https://github.com/PolyChord/PolyChordLite}}, using 250 live points and uniform priors as listed in table \ref{T:prior}. We set a precision criterion of 0.001 as the convergence threshold. The resulting output samples are analyzed using \texttt{GetDist} to obtain the parameter constraints. 
%%%%%%%%%%%%%%%%%%%%%%%%%%%%%%%%%
\begin{table}[h]
    \begin{center}
        \begin{tabular}{|p{0.19\textwidth} |p{0.19\textwidth}|p{0.19\textwidth}|p{0.19\textwidth}|}
        \hline
        Method&  $\Omega_m$ & $h$ & $M$\\
        & & &  \\
        \hline
        GLS  & $0.306 \pm 0.010$ & $0.722 \pm 0.983 $& $-19.217 \pm 2.96$\\
        & &  & \\
        \hline
        Jackknife delete-1 & $ 0.306\pm 0.009\;\;$ ($0.0001$)& $ 0.696 \pm 0.002\;\;$ ($0.025$)& $-19.293 \pm 0.014\;\;$ ($0.076$)\\
        & &  & \\
        \hline
        Jackknife delete-41 & $ 0.303 \pm 0.009\;\;$ ($0.004$)& $0.707 \pm 0.006\;\;$ ($0.015$)& $ -19.262\pm 0.025\;\;$ ($0.045$)\\
        & &  & \\
        \hline
        Jackknife delete-170 & $ 0.305\pm 0.009\;\;$ ($0.001$) & $0.715 \pm 0.004\;\;$ ($0.0067$) & $-19.239 \pm 0.020\;\;$ ($0.022$)\\
        &  & & \\
        \hline
        Bootstrap & $0.306 \pm 0.014\;\;$ ($0.00008$) & $0.721 \pm 0.002 \;\;$ ($0.0002$) & $ -19.218\pm 0.016\;\;$ ($0.00079$)\\
        &  & & \\
        \hline
        Nested Sampling & $ 0.306\pm 0.010$ & $0.737 \pm 0.104$ & $ -19.193\pm 0.312$\\
        & & & \\
        \hline
        \end{tabular}
        \caption{\label{mock_covariance_corrected}The table lists the parameter estimates and associated uncertainties obtained for the mock flat $\Lambda$CDM dataset using all methods. Note that Jackknife detects a bias in GLS estimates of $h$ and $M$. The bias-corrected estimates are closer to the fiducial values, with tighter error bars. 
        }
    \end{center}
\end{table}
%%%%%%%%%%%%%%%%%%%%%%%%%%%%%%%%%%%%%%%%%%%%%%%%%
\section{Calibration of Techniques}\label{sec:calibration}
Before we apply the methods described above to the PPS data set, we calibrate them. We perform two tests. Firstly, we apply methods to a near-ideal mock data set. Secondly, we study the effect of data quality on the different statistical methods. 
%%%%%%%%%%%%%%%%%%

\subsection{Calibration using a mock data set}\label{sub:mock}
For the mock data, we consider a fiducial flat $\Lambda$CDM model with parameters set to $\Omega_m\,=\,0.3$, $h\,=\,0.7$ and $M\,=\,-19.3$. We generate a sample of $1701$ number of supernovae equally spaced in redshift over the range $0.001<z<2.261$. We assume a one per cent Gaussian-distributed error in the data. Unlike PantheonPlus data, we do not consider any covariance between different data points. Since Bayesian methods are standard and well established, we focus on frequentist techniques. We, however, present the analysis using nested sampling for comparison. Table \ref{mock_covariance_corrected} lists the estimates of best-fit values and their standard deviation obtained using GLS, Jackknife, and Bootstrap methods. The best-fit estimates quoted for Jackknife and Bootstrap are bias-corrected, with the bias shown in parentheses. The table also lists the marginalised mean values and their associated uncertainties obtained using the Bayesian approach with the nested sampling algorithm.
%%%%%%%%%%%%%%%%%%%%%%%%%%%%%%%%%%
\begin{table}[h]
    \begin{center}
        \begin{tabular}{|p{0.19\textwidth} |p{0.19\textwidth}|p{0.19\textwidth}|p{0.19\textwidth}|}
        \hline
        Method&  $\Omega_m$ & $h$ & $M$\\
        & & &  \\
        \hline
        GLS  & $0.306 \pm 0.010$ & $0.695 \pm 0.004 $& $-19.3$\\
        & &  & \\
        \hline
        Jackknife delete-1 & $ 0.306\pm 0.009\;\;$ ($0.0001$)& $ 0.694 \pm 0.004\;\;$ ($-0.0000023$)& $-19.3$\\
        & &  & \\
        \hline
        Jackknife delete-41 & $ 0.303 \pm 0.009\;\;$ ($0.004$)& $0.696 \pm 0.004\;\;$ ($-0.001$)& $ -19.3$\\
        & &  & \\
        \hline
        Jackknife delete-170 & $ 0.306\pm 0.009\;\;$ ($0.0001$) & $0.695 \pm 0.004\;\;$ ($0.00002$) & $-19.3$\\
        &  & & \\
        \hline
        Bootstrap & $0.306 \pm 0.015\;\;$ ($-0.0002$) & $0.695 \pm 0.006 \;\;$ ($0.00008$) & $ -19.3$\\
        &  & & \\
        \hline
        Nested Sampling & $ 0.306\pm 0.010$ & $0.694 \pm 0.004$ & $ -19.3$\\
        & & & \\
        \hline
        \end{tabular}
        \caption{\label{mock_covariance_corrected_fixM}Same as table \ref{mock_covariance_corrected}, but with $M$ set to a fixed value. We find that breaking the degeneracy between $h$ and $M$ makes estimates obtained with all methods similar. The estimates and their error bars are similar to those obtained with Jackknife delete-1 in table \ref{mock_covariance_corrected}, wherein degeneracy was not broken.}
    \end{center}
\end{table}
%%%%%%%%%%%%%%%%%%%%%%%%%%%%%%%%%%%%%%%
\par 
We find that estimates of $\Omega_m$ and its error bar are similar across all methods. Estimates for $h$ and $M$ obtained with GLS and Bayesian methods are comparable, with GLS yielding a larger error estimate. Compared to the error in estimating $\Omega_m$, the order of error of $h$ and $M$ is considerably larger. Jackknife analysis reveals bias in the GLS estimates of $h$ and $M$, and we find that the bias-corrected estimates are significantly closer to the fiducial values. Jackknife also puts a tighter error bar on its estimate. Increasing the number of delete points decreases the measured bias. We note that though Jackknife delete-1 and delete-41 estimates of $h$ and $M$ are within 1-2 $\sigma$, delete-170 estimates $h$ and $M$ to be over $3 \sigma$ away from the fiducial value. So, increasing delete points beyond a certain value is not advised. The error estimated by Bootstrap is comparable to that obtained with Jackknife. However, it fails to estimate bias, causing the Bootstrap estimates of $h$ and $M$ to be about $11 \sigma$ and $5.12 \sigma$ away from the fiducial value, respectively. 
\par 
Table \ref{mock_covariance_corrected_fixM} reports the estimates of parameters when the degeneracy between $h$ and $M$ is broken by fixing the value of $M$. We note that estimates of $\Omega_m$ is same as that obtained in table \ref{mock_covariance_corrected}. However, the estimates of $h$
 becomes similar to the Jackknife delete-1 estimate obtained in table \ref{mock_covariance_corrected}. This further illustrates that, for the near-ideal data set, the estimate of the Jackknife was better in the presence of correlation between $h$ and $M$. 

\par 
%%%%%%%%%%%%%%%%%%
\begin{table}
    \begin{center}
        \begin{tabular}{|p{0.17\textwidth}|p{0.17\textwidth} |p{0.17\textwidth}|p{0.17\textwidth}|p{0.17\textwidth}|}
        \hline
        Method&  Data & $\Omega_m$ & $h$ & $M$\\
        & & & & \\
        \hline
        & PantheonPlus & $0.361 \pm 0.019$ & $0.740 \pm 0.636$ & $-19.220 \pm 1.87$\\
        & & & & \\
        GLS & PantheonPlus ($z>0.01$) &$0.332 \pm 0.018$ &$0.727 \pm 0.029$ & $-19.271 \pm 0.087$\\
        & & & & \\
        & PantheonPlus and SH0ES & $0.333\pm  0.018$& $0.735 \pm 0.010$& $-19.248 \pm 0.029$\\
        & & & & \\
        \hline
        & PantheonPlus &  $0.323 \pm 0.027\;\;$ ($0.038$) &  $9.648 \pm 0.1\;\;\;$\hspace{1cm}  ($-8.91$)  & $7.01 \pm 0.282\;\;$ ($-26.2$)\\
        & & & & \\
        Jackknife delete-1 & PantheonPlus ($z>0.01$) & $0.287 \pm 0.016\;\;$ ($0.044$)& $0.678 \pm  0.052\;\;$  ($ 0.049$)& $-19.429 \pm  0.153\;\;$ ($0.158$)\\
        & & & & \\
        & PantheonPlus and SH0ES & $0.297 \pm 0.016\;\;$ ($0.036$)& $0.743 \pm 0.029\;\;$ ($-0.008$)& $ -19.237 \pm 0.087\;\;$ ($-0.011$)\\
        & & & & \\
        \hline
        & PantheonPlus & $0.351 \pm 0.030\;\;$  ($0.010$) & $0.743 \pm 0.005\;\;$ ($  -0.003$) & $-19.217 \pm 0.014\;\;$ ($-0.003$)\\
        & & & & \\
        Bootstrap & PantheonPlus ($z>0.01$) & $0.317 \pm 0.028\;\;$ ($0.015$)& $0.711 \pm  0.026\;\;$ ($0.016$) & $-19.323 \pm  0.079\;\;$ ($0.052$) \\
        & & & & \\
        & PantheonPlus and SH0ES & $0.320 \pm 0.026\;\;$ ($0.013$)& $0.736 \pm 0.011\;\;$ ($-0.0018$)& $-19.248 \pm 0.031\;\;$ ($0.00004$)\\
        & & & & \\
        \hline
        & PantheonPlus &  $0.362 \pm  0.019$ & $0.725 \pm 0.105$ & $-19.289 \pm 0.317$\\
        & & & & \\
        Nested sampling & PantheonPlus ($z>0.01$) & $0.332 \pm 0.018$ & $0.729 \pm 0.030$ & $-19.266 \pm 0.090$\\
        & & & & \\
        & PantheonPlus and SH0ES & $0.333 \pm 0.018$ & $0.735 \pm 0.010$ & $-19.247 \pm 0.029$\\
        & & & & \\
        \hline
        \end{tabular}
        \caption{\label{T:diffdata}The table lists the constraints on the flat $\Lambda$CDM parameters obtained from three different versions of PantheonPlus datasets. As in table \ref{mock_covariance_corrected}, the estimates of best fit parameters obtained with Jackknife and Bootstrap resampling techniques are bias corrected, with the estimate of bias given in parentheses. 
        The Jackknife estimate of $h$ from PP[$z>0.01$] is fully consistent with the Planck estimate. 
        From PPS, Jackknife, and Bootstrap derive a bias in the estimate of $\Omega_m$ and hence their estimates are smaller.  Estimates of $h$ and $M$ are similar across methods, though Jackknife estimates a higher error bar.  
}
    \end{center}
\end{table}
%%%%%%%%%%%%%%%%%%
\subsection{Calibration with different datasets}\label{sub:diff_data}
In this subsection, we analyze the flat $\Lambda$CDM model with three different versions of PantheonPlus data: (i) PantheonPlus and SH$0$ES (PPS) with $N = 1701$ data points, (ii) PantheonPlus (PP) with $N = 1701$ data points, (iii) PantheonPlus with only $z>0.01$ (PP[$z>0.01$]), which has $N= 1590$ data points. 
Since the inclusion of SH0ES partly breaks the degeneracy between $h$ and $M$, we expect PPS to be the cleanest data. Further, since low-redshift data is plagued by systematic errors due to peculiar velocities \cite{Scolnic:2021amr}, restricting the data to $z>0.01$ also makes it cleaner. 
Within this analysis, we consider two aspects, namely: (1) the comparison of estimates from different datasets, and (2) the variation in Jackknife estimates with the number of delete points $d$. The results obtained are summarized in the tables \ref{T:diffdata} and \ref{T:jackknife_comp}.
%%%%%%%%%%%%%%%%%%%%%%%%%%%
\subsubsection{Comparison of estimates with different datasets}
In this section, we assess how parameter constraints vary across versions of the PantheonPlus data. From table \ref{T:diffdata}, we note the following. (1) Estimates of $\Omega_m$ are similar across datasets for GLS and Bayesian methods. All methods report higher value of $\Omega_m$ with PP. Jackknife reports a bias in estimate of $\Omega_m$ and hence its bias-corrected estimates are lowest across all datasets. (2) Estimates of $h$ and $M$ are largely comparable across datasets for GLS and Bayesian methods, with GLS reporting a larger error. (3) Jackknife delete-1 estimates atypical values for $h$ and $M$ with a very large bias. For PP [$z > 0.01$], the estimate of $h$ is lower and seems to be consistent with estimates from Planck \cite{Planck:2018vyg}. Here, we note that bias is comparable to standard deviation. Estimates obtained with PPS are consistent with other methods. However, we note that the Jackknife reports larger error bars for $h$ and $M$. (4) The Bootstrap estimates lower bias than the Jackknife. With PPS, Bootstrap estimates are largely comparable to Bayesian, with some departure in the case of $\Omega_m$ due to bias as well as a larger error bar.
%%%%%%%%%%%%%%%%%%%%%%%%%%%%%%%%
\subsubsection{Variation in Jackknife estimates with the number of delete points $d$.}\label{subsub:dvariation}
Parameter estimates obtained with different datasets for $d = 1$, $d = 41$ and $d = 170$ are given in table \ref{T:jackknife_comp}. 
From this table, we find the following. (1) The estimates of $\Omega_m$ are largely independent of the value of $d$ across all datasets. (2) For the PP data, the bias in $h$ and $M$ decreases with an increase in $d$. However, since error also decreases with $d$, the estimates of $h$ and $M$ do not become fully consistent with the estimates from other methods. (3) Interestingly, we note that the estimate of $h$ from PP[$z>0.01$] is fully consistent with the Planck estimate. This result holds to be true regardless of the value of $d$. 
%%%%%%%%%%%%%%%%
\begin{table}
    \begin{center}
        \begin{tabular}{|p{0.19\textwidth}|p{0.17\textwidth} |p{0.17\textwidth}|p{0.17\textwidth}|p{0.17\textwidth}|}
        \hline
        Method&  Data & $\Omega_m$ & $h$ & $M$\\
       & & & & \\
        \hline
        & PantheonPlus &  $0.323 \pm 0.027\;\;$ ($0.038$) &  $9.648 \pm 0.095\;\;\;$\hspace{1cm}  ($-8.91$)  & $7.01 \pm 0.282\;\;$ ($-26.2$)\\
        & & & & \\
        Jackknife delete-1& PantheonPlus ($z>0.01$) & $0.287 \pm 0.016\;\;$ ($0.044$)& $0.678 \pm  0.052\;\;$  ($ 0.049$)& $-19.429 \pm  0.153\;\;$ ($0.158$)\\
        & & & & \\
        & PantheonPlus and SH0ES & $0.297 \pm 0.016\;\;$ ($0.036$)& $0.743 \pm 0.029\;\;$ ($-0.008$)& $ -19.237 \pm 0.087\;\;$ ($-0.011$)\\
        & & & & \\
        \hline
        & PantheonPlus &  $0.322\pm0.025 \;\;$ ($0.039$) &  $ 0.931\pm 0.017\;\;\;$\hspace{1cm}  ($-0.191$)  & $ -18.677\pm 0.051\;\;$ ($-0.543$)\\
        & & & & \\
        Jackknife delete-41& PantheonPlus ($z>0.01$) & $ 0.291\pm 0.016\;\;$ ($0.040$)& $ 0.691 \pm 0.045 \;\;$  ($ 0.036$)& $ -19.390\pm 0.133 \;\;$ ($0.119$)\\
        & & & & \\
        & PantheonPlus and SH0ES & $ 0.303\pm0.016 \;\;$ ($0.030$)& $ 0.736\pm 0.027\;\;$ ($-0.001$)& $  -19.256\pm 0.080 \;\;$ ($0.008$)\\
        & & & & \\
        \hline
        & PantheonPlus &  $ 0.325\pm 0.023\;\;$ ($0.035$) &  $ 0.774\pm 0.007\;\;\;$\hspace{1cm}  ($-0.034$)  & $ -19.135 \pm 0.023\;\;$ ($-0.085$)\\
        & & & & \\
        Jackknife delete-170 & PantheonPlus ($z>0.01$) & $ 0.296\pm 0.015\;\;$ ($0.035$)& $ 0.683\pm 0.039 \;\;$  ($0.044$)& $ -19.411\pm 0.114\;\;$ ($0.140$)\\
        & & & & \\
        & PantheonPlus and SH0ES & $ 0.306 \pm 0.015\;\;$ ($0.027$)& $ 0.741\pm 0.017\;$ ($-0.007$)& $  -19.240\pm 0.050\;\;$ ($-0.008$)\\
        & & & & \\
        \hline
        \end{tabular}
        \caption{\label{T:jackknife_comp} Same as table \ref{T:diffdata} but for Jackknife with different delete points. We find that estimates of $\Omega_m$ are largely independent of the  $d$ points. For $h$ and $M$, this is only true with PP[$z > 0.01$] and PPS. Interestingly, we note that estimate of $h$ obtained from PP[$z > 0.01$] is fully consistent with Planck estimate. This is true regardless of the value of $d$. 
 }
    \end{center}
\end{table}
%%%%%%%%%%%%%%%%%%%%%%%%%%
%%%%%%%%%%%%%%%%%%%%%%%
\subsection{Conclusions from calibration}
From the analyses described in the above two subsections (\ref{sub:mock}, \ref{sub:diff_data}), we conclude the following:
\begin{itemize}
\item With the near-ideal mock data, Jackknife is the most successful in recovering the fiducial values, since its estimates are the closest to the fiducial values with small error. 
\item Across datasets, the best-fit frequentist GLS estimates are found to be largely consistent with the Bayesian estimates. The difference lies in the larger uncertainty estimated for the parameters $h$ and $M$ in the GLS method. 
    \item As expected, the estimates from both Jackknife and Bootstrap resampling techniques are highly dependent on the dataset. The Bootstrap method generally underestimates the bias. We find that, as the data gets better, for instance with PP[$z>0.01$] and PPS, the agreement between different methods also improves. 
    \item From the analysis of mock data, we find that Jackknife delete-1 and delete-41 perform well. Further, their estimates from PPS data are comparable, with Jackknife delete-41 estimating a lower bias. Hence, we report constraints obtained with both Jackknife delete-1 and delete-41 for the other parameterisations of dark energy. 
\end{itemize}
%%%%%%%%%%%%%%%%%
\begin{table}
\begin{center}
    \begin{tabular}{|p{0.12\textwidth}|p{0.19\textwidth}|p{0.18\textwidth}|p{0.19\textwidth}|p{0.19\textwidth}|}
    \hline
    Model&Method&$\Omega_m$&$h$&$M$\\
    &&&&\\
    \hline
    & GLS & $0.333\pm  0.018$ & $0.735 \pm 0.010$ & $-19.248 \pm 0.029$\\
   &&&&\\
   & Jackknife delete-$1$ & $0.297 \pm 0.016$ \;\; ($0.036$) & $ 0.743 \pm 0.029\;\;$ ($-0.008$) & $ -19.237 \pm 0.087$ \;\; ($-0.011$)\\
    &&&&\\
    Flat $\Lambda$CDM& Jackknife delete-41& $ 0.303 \pm 0.016$ \;\; ($0.030$)& $ 0.736\pm 0.027 \;\;$ ($-0.001$)& $ -19.256\pm 0.080$ \;\; ($0.008$)\\
    &&&&\\
    &Bootstrap & $0.320 \pm 0.026\;\;$ ($0.013$ ) & $0.736 \pm 0.011\;\;$ ($-0.0018$) & $-19.248 \pm 0.031$\;\; ($0.00004$)\\
    &&&&\\
    &MCMC& $0.333 \pm 0.018$ & $0.735 \pm 0.010$ & $-19.248 \pm 0.029$ \\
    &&&&\\
    & Nested sampling& $0.333 \pm 0.018$ & $0.735 \pm 0.010$ &$-19.247 \pm 0.029$\\
    &&&&\\
    \hline
    \end{tabular}\caption{\label{T:Flatlcdm}The table lists the estimates for the flat $\Lambda$CDM model obtained using the PPS dataset with both the frequentist and Bayesian methods. The quoted mean estimates against the resampling techniques, Jackknife and Bootstrap methods, are bias-corrected estimates, with bias shown in parentheses. We find that both the Bayesian approaches lead to largely identical constraints, which also show agreement with the GLS estimates. Both Jackknife and Bootstrap detect bias in estimate of $\Omega_m$ leading to a lower value. Nevertheless, the bias corrected estimates are consistent with other methods within 1-2 $\sigma$. We also note that estimate of $h$ and $M$ obtained with Jackknife are within $1\sigma$ of estimates from other methods.  
    These results indicate that, for the simplest three-parameter flat $\Lambda$CDM model, all methods yield consistent results. Interestingly, both Jackknife methods estimates a larger error for $h$ and $M$. The corresponding corner plot obtained with all methods is shown in figure \ref{fig:flcdm_delete1}.}
\end{center}
\end{table} 
%%%%%%%%%%%%%%%%%%%%%%%%%%%%%%%%%%%%%%%%%%%%%%%%%%%%%%%%%%%%%%%%%%%%%%%%%
\begin{figure}
	\centering
	\begin{tabular}{c}
		\includegraphics[width=0.5\linewidth]{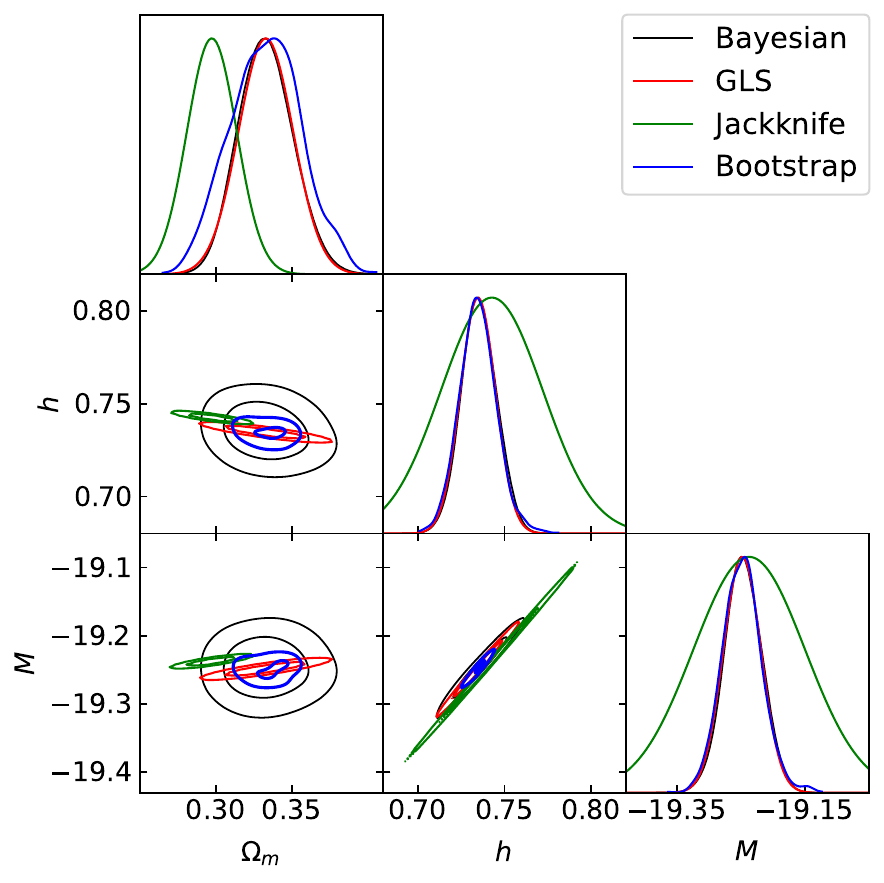}
	\end{tabular}
	 \caption{\label{fig:flcdm_delete1} Corner plots for flat $\Lambda$CDM. Jackknife refers to Jackknife delete-1. See table \ref{T:Flatlcdm} for estimates of mean, error, bias and a discussion. 
 }
\end{figure}
%%%%%%%%%%%%%%
\begin{table}
\begin{center}
    \begin{tabular}{|p{0.1\textwidth}|p{0.19\textwidth}|p{0.14\textwidth}|p{0.14\textwidth}|p{0.14\textwidth}|p{0.15\textwidth}|}
    \hline
     Model&Method&$\Omega_m$&$\Omega_{\rm DE}$&$h$&$M$\\
    &&&&&\\
    \hline
    & GLS & $0.297 \pm 0.054$ & $0.612 \pm 0.080$ & $0.734 \pm 0.010$  & $-19.248 \pm 0.029$\\
    &&&&&\\
    & Jackknife delete-$1$& $0.289 \pm 0.041 \;\; $ ($0.007$) & $0.679 \pm 0.070$ \;\;($-0.067$) &  $0.744 \pm 0.029  \;\;$ ($–0.010$) & $-19.232 \pm 0.087$ \;\;($-0.016$)\\
    &&&&&\\
    $\Lambda$CDM &Jackknife delete-41& $ 0.279\pm 0.042$ \;\; ($ 0.017$)& $0.672 \pm 0.067$ \;\; $(-0.060)$ & $ 0.739\pm0.027 $ \;\;($-0.005$)& $ -19.247\pm 0.080$ \;\; ($-0.0002 $)\\
    &&&&&\\
    & Bootstrap& $0.293 \pm 0.072\;\;$ ($0.003$)& $0.632 \pm 0.109$ ($-0.020$) &$0.735 \pm 0.012$ \;\; ($-0.001$)& $-19.249 \pm 0.032$\;\; ($0.0009$) \\
    &&&&&\\
    &MCMC& $0.296 \pm 0.054$ & $0.609 \pm 0.080$ &  $0.734 \pm 0.010$  & $-19.247 \pm 0.029$\\
    &&&&&\\
    & Nested sampling& $0.297 \pm 0.054$ & $0.611 \pm 0.082$ & $0.734 \pm 0.010$ & $-19.248 \pm 0.029$  \\
     &&&&&\\
    \hline
    \end{tabular}\caption{\label{T:lcdm}Same as table \ref{T:Flatlcdm}, but for the $\Lambda$CDM model. All methods give consistent results within $1\sigma$. Jackknife estimates a higher error bar for $h$ and $M$. The bias correction pushes the estimates of $\Omega_{\rm DE}$ (especially from Jackknife) to larger values, indicating that the Universe is flat. The corresponding corner plot for the $\Lambda$CDM model is shown in the left panel of figure \ref{fig:lcdm_fwcdm_delete1}.}
    \end{center}
    \end{table} 

%%%%%%%%%%%%%%%%%%%%%%%%%%%
\begin{table}
\begin{center}
    \begin{tabular}{|p{0.12\textwidth}|p{0.18\textwidth}|p{0.14\textwidth}|p{0.14\textwidth}|p{0.14\textwidth}|p{0.145\textwidth}|}
    \hline
    Model&Method&$\Omega_m$&$w_0$&$h$&$M$\\
    &&&&&\\
        \hline
    & GLS & $0.288 \pm 0.071$ & $-0.891 \pm 0.148$  &$0.734 \pm 0.010$ & $-19.248\pm0.029$ \\
    &&&&&\\
    & Jackknife delete-$1$& $0.288 \pm0.057$ \;\; ($-0.0002$)& $-0.942\pm0.128$\;\; ($0.051$) & $0.743 \pm 0.029$ \;\; ($-0.010$) & $-19.231\pm0.087$ \;\; ($-0.016$)  \\
    &&&&&\\
    Flat $w$CDM & Jackknife delete-41& $ 0.292\pm 0.059$ \;\; ($-0.004$)& $ -0.943\pm  0.131$ \;\; ($0.053 $) & $ 0.749 \pm  0.025$ \;\; ($ -0.016$) & $-19.216\pm 0.074$ \;\; ($ -0.032$)  \\
    &&&&&\\
    & Bootstrap& $ 0.294 \pm 0.101$ \;\; ($ -0.006$ )& $-0.880 \pm 0.213$\;\; ($-0.010$)& $0.736 \pm 0.012$\;\; ($-0.002 $)&$-19.246\pm0.032$ \;\; ($-0.001$)  \\
    &&&&&\\
    &MCMC&$0.286 \pm 0.071$ &  $-0.908 \pm 0.149$& $0.734 \pm 0.010$& $-19.247 \pm 0.029$ \\
    &&&&&\\
    & Nested sampling& $0.290 \pm 0.069$& $-0.913 \pm 0.146$ &  $0.734 \pm 0.010$& $-19.247 \pm 0.029$\\
    &&&&&\\
    \hline
    \end{tabular}\caption{\label{T:fwcdm}Same as table \ref{T:Flatlcdm}, but for the flat $w$CDM model. We find that all the methods yield largely similar constraints, within $1\sigma$, on all four free parameters. Estimates are consistent with dark energy being a cosmological constant. The bias correction makes the Jackknife estimate of $w_0$ closer to $-1$ than other methods. Jackknife estimates higher error bars for $h$ and $M$. We also note that, unlike other models, bias to error ratio is small for all parameters. The corresponding corner plot obtained with all methods is shown in the right panel of figure \ref{fig:lcdm_fwcdm_delete1}.}
    \end{center}
    \end{table}
%%%%%%%%%%%%%%%%%%%%%%%%%%%%%%%%%%%%%%%%%%%%%%%%%%%%%%%
\begin{figure}
	\centering
	\begin{tabular}{cc}
		\includegraphics[width=0.5\linewidth]{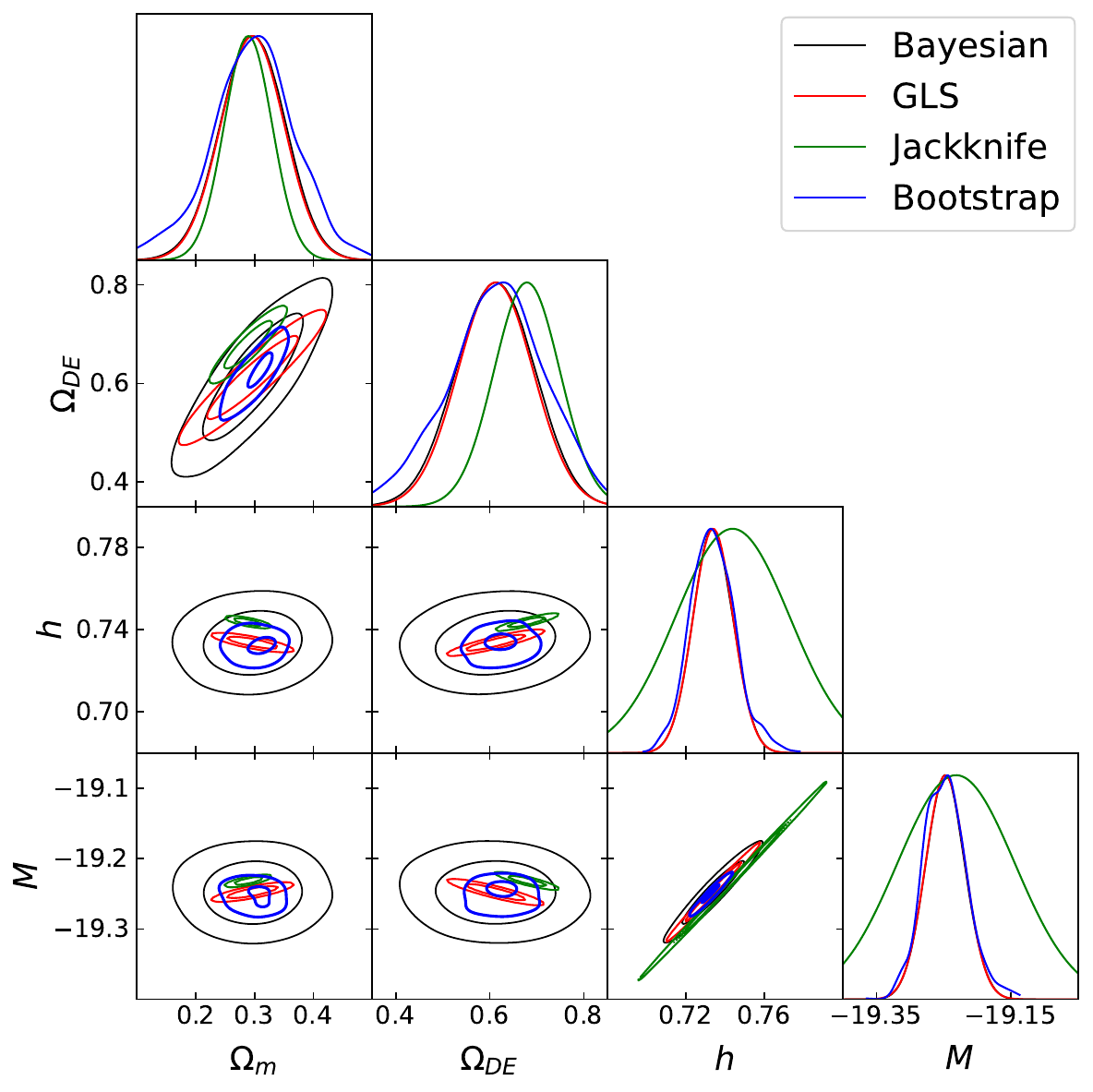}&
		\includegraphics[width=0.5\linewidth]{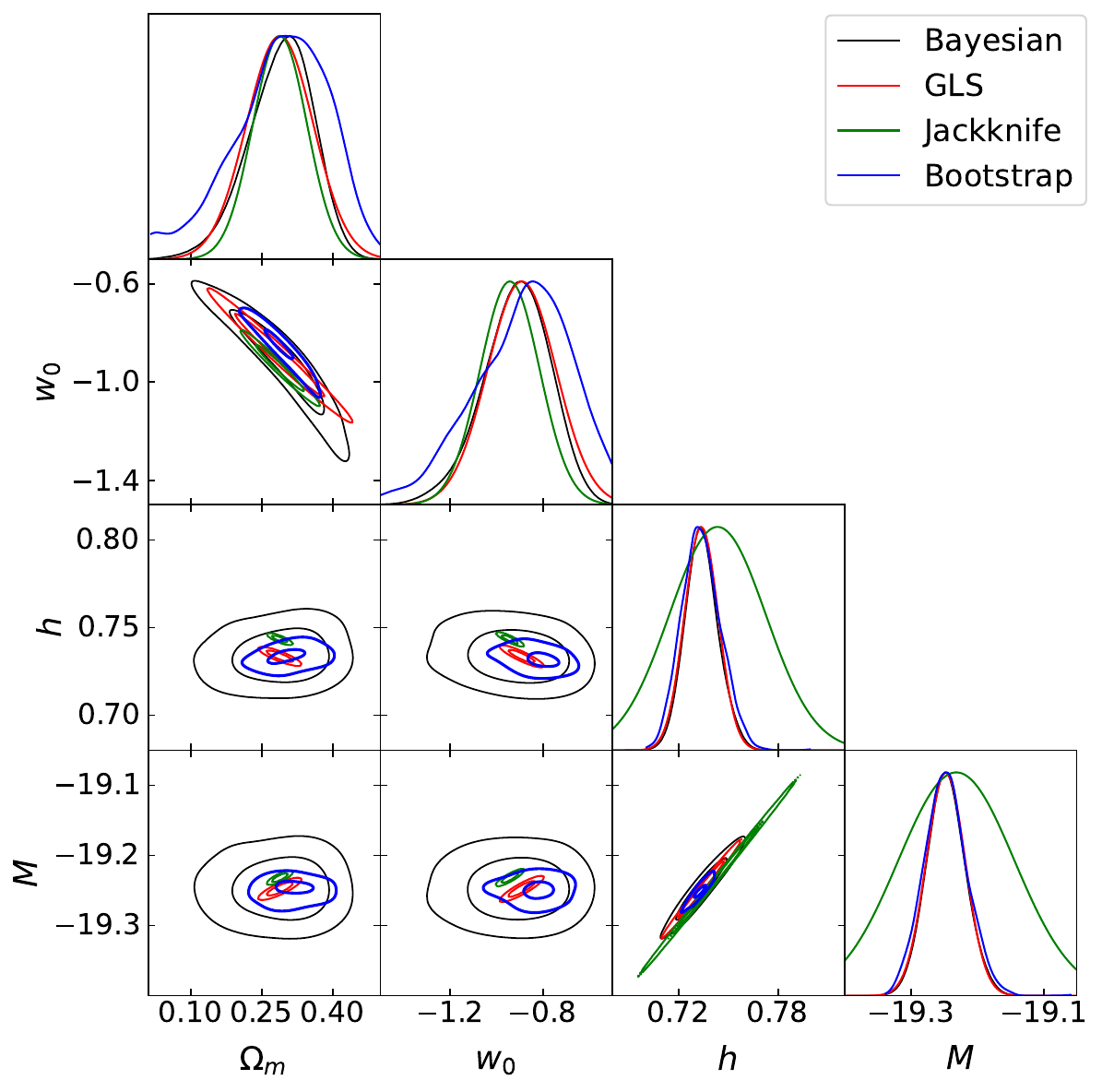}
	\end{tabular}
	\caption{\label{fig:lcdm_fwcdm_delete1} Same as Figure \ref{fig:flcdm_delete1}, but for $\Lambda$CDM (left) and flat $w$CDM (right) models. See table \ref{T:lcdm} and \ref{T:fwcdm} for estimates of mean, error, bias and a discussion.
    }
\end{figure}
%%%%%%%%%%%%%%%%%%%%%%%%%%%%%
\begin{table}
\begin{center}
    \begin{tabular}{|p{0.08\textwidth}|p{0.16\textwidth}|p{0.14\textwidth}|p{0.14\textwidth}|p{0.14\textwidth}|p{0.14\textwidth}|p{0.15\textwidth}|}
    \hline
    Model&Method&$\Omega_m$&$\Omega_{\rm DE}$&$w_0$&$h$&$M$\\
    &&&&&&\\
    \hline
    &GLS &$0.296 \pm 0.074$ & $0.625 \pm 0.762$ & $-0.983 \pm 0.935$ & $0.734 \pm 0.010$ &  $-19.248 \pm 0.029$\\
   &&&&&&\\
    &Jackknife delete-$1$& $ 0.479 \pm 0.073\;\;$ ($-0.182$)& $0.205 \pm 0.605$ \;\;($0.421$) &  $-0.796 \pm 0.779$ \;\;($-0.187$) & $0.743 \pm 0.029 $ \;\; ($-0.009$ )& $-19.232 \pm 0.087 $ \;\; ($-0.016$)\\
    &&&&&&\\
    $w$CDM &Jackknife delete-41& $ 0.668\pm0.266 \;\;$ ($-0.372 $)& $ -0.441\pm 1.03$ \;\;($1.066 $) &  $ -1.08\pm0.833 $ \;\;($ 0.096$) & $ 0.745\pm 0.026$ \;\; ($-0.0114$ )& $ -19.224\pm 0.078 $ \;\; ($ -0.023$)\\
    &&&&&&\\
    &Bootstrap& $0.392 \pm 0.165$ \;\; ($-0.096$)& $0.542 \pm 0.524$ \;\; ($0.084$)& $-0.463 \pm 1.135$ \;\; ($-0.520$)& $0.735 \pm 0.012$ \;\; ($-0.0017$)& $-19.244 \pm 0.032$ \;\; ($-0.0033$)\\
    &&&&&&\\
    &MCMC & $0.241 \pm 0.082$& $0.697 \pm 0.364$ &  $-1.156 \pm 0.580$ & $0.734 \pm 0.010$ & $-19.247 \pm 0.030$ \\
    &&&&&&\\
    & Nested sampling& $0.251 \pm 0.072$ & $0.574 \pm 0.328$ & $-1.403 \pm 0.734$ & $0.735 \pm 0.011$ & $-19.247 \pm 0.030$\\
    &&&&&&\\
    \hline
   \end{tabular}\caption{\label{T:wcdm}Same as table \ref{T:Flatlcdm}, but for the five-parameter $w$CDM model. The GLS and Bayesian methods still provide broadly consistent constraints for all the parameters. Resampling techniques estimate atypical values for $\Omega_m$, $\Omega_{\rm DE}$, and $w_0$ due to greater bias. Estimates of $h$ and $M$ of all methods are consistent with each other, though Jackknife reports a larger error bar. In general, we also note that PPS data are insufficient to constrain this five-parameter model. The corresponding corner plot obtained with all methods is shown in the left panel of figure \ref{fig:wcdm_fw0wacdm_delete1}.}
\end{center}
\end{table}

\begin{table}
\begin{center}
     \begin{tabular}{|p{0.102\textwidth}|p{0.16\textwidth}|p{0.14\textwidth}|p{0.14\textwidth}|p{0.14\textwidth}|p{0.14\textwidth}|p{0.145\textwidth}|}
    \hline
    Model&Method&$\Omega_m$&$w_0$&$w_a$&$h$&$M$\\
    &&&&&&\\
    \hline
    & GLS & $0.266 \pm 0.313$ & $-0.874 \pm 0.290 $ & $0.174 \pm 2.120$& $0.734 \pm 0.010$ & $-19.248 \pm 0.029$ \\
    &&&&&&\\
    & Jackknife delete-1&$0.693\pm0.262$\;\; ($-0.427$) & $-1.357 \pm 0.236$ \;\; ($0.483$)& $-0.503 \pm 1.573$ \;\; ($0.676$)& $0.743 \pm 0.029$ \;\; ($-0.010$)& $-19.232 \pm 0.087$  \;\; ($-0.016$) \\
  &&&&&&\\
     Flat $w_0\,w_a$CDM & Jackknife delete-$41$ &$ 0.749 \pm 0.270\;\;$ ($-0.482$) & $  -1.413 \pm  0.231 $ \;\; ($ 0.539$)& $ -0.699 \pm 1.49 $ \;\; ($ 0.873$)& $0.759 \pm 0.024$ \;\; ($-0.025 $)& $ -19.187 \pm 0.072$  \;\; ($-0.060$) \\
    &&&&&&\\
     & Bootstrap&$0.275 \pm 0.145$\;\; ($-0.009$) & $-0.917 \pm 0.199$ \;\; ($0.043$) & $0.942 \pm 1.968$ \;\; ($-0.769$)& $0.736 \pm 0.012$\;\; ($-0.003$) & $-19.246 \pm 0.032$ \;\; ($-0.002$) \\
     &&&&&&\\
    &MCMC&$0.322 \pm 0.106$ &$-0.919 \pm 0.148$ & $-0.826 \pm 1.35$& $0.733 \pm 0.010 $& $-19.247 \pm 0.030$ \\
    &&&&&&\\
    & Nested sampling& $0.315 \pm 0.114$&$-0.913 \pm 0.148$ & $-0.816 \pm 1.40$ & $0.733 \pm 0.010$ & $-19.248 \pm 0.030$\\
    &&&&&&\\
    \hline
    \end{tabular}\caption{\label{T:fw0wacdm}Same as table \ref{T:Flatlcdm} model, but for the five parameter flat $w_0\,w_a$CDM model. The GLS and Bayesian methods still provide broadly consistent constraints for all the parameters, with GLS estimating larger errors. Resampling techniques estimate atypical values for $\Omega_m$, $w_0$, and $w_a$ due to greater bias. Estimates of $h$ and $M$ of all methods are consistent with each other, though Jackknife reports a larger error bar. In general, we also note that PPS data are insufficient to constrain this five-parameter model. The corresponding corner plot obtained with all methods is shown in the right panel of figure \ref{fig:wcdm_fw0wacdm_delete1}.
    }
\end{center}
\end{table}

%%%%%%%%%%%%%%%%%%%%%%%
%%%%%%%%%%%%%%%%%%%%%%%
\begin{figure}
\begin{tabular}{cc}
		\includegraphics[width=0.5\linewidth]{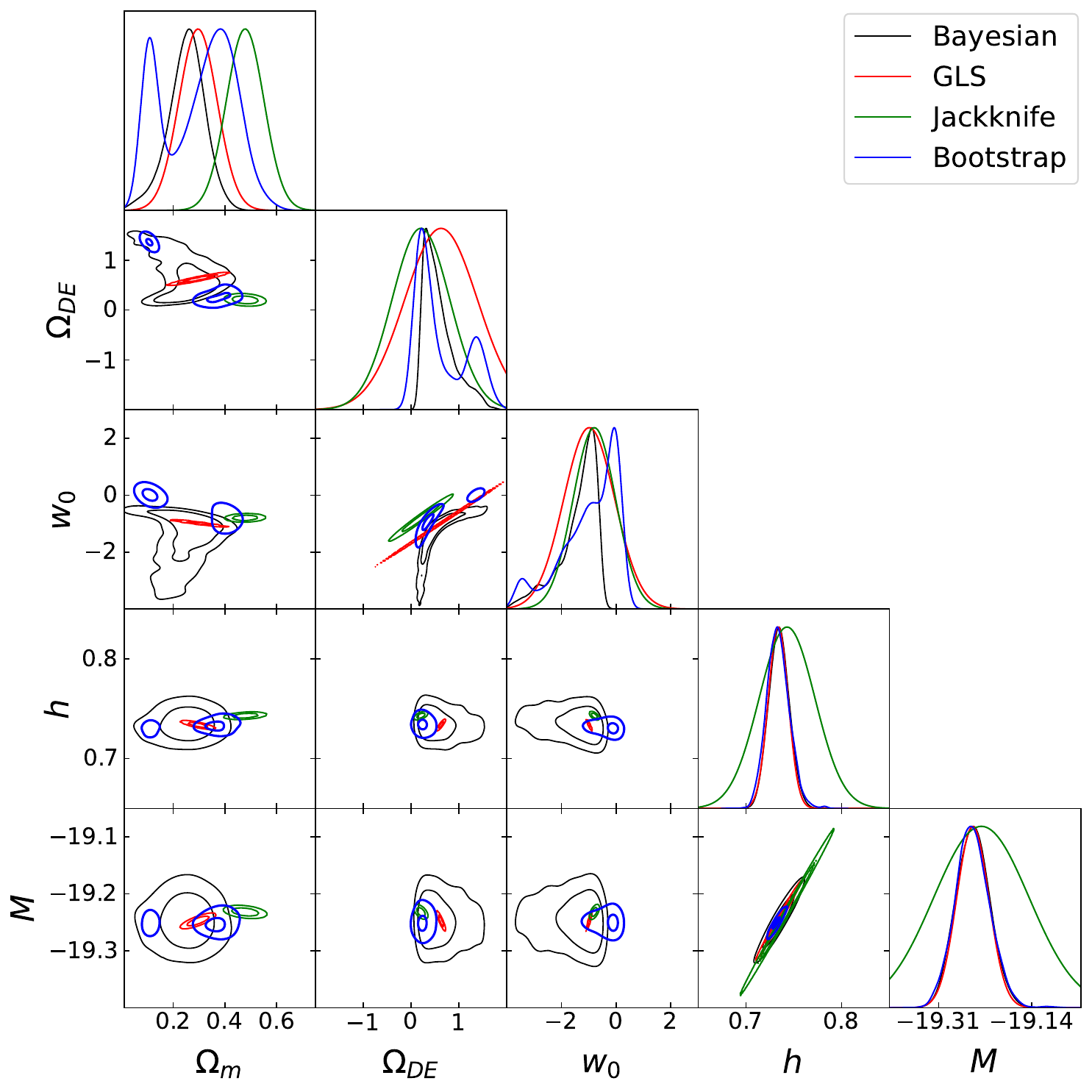}&
		\includegraphics[width=0.5\linewidth]{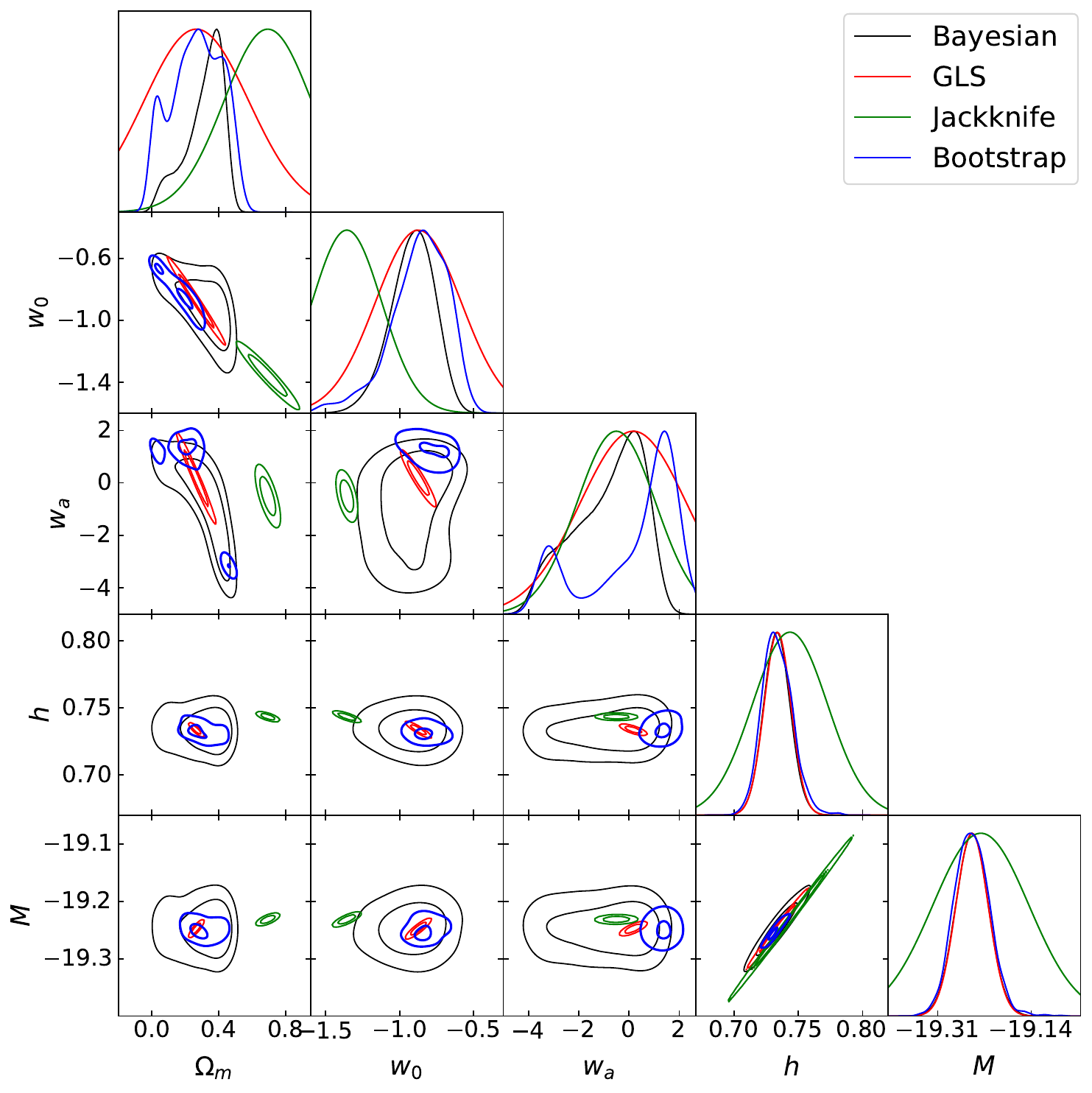}
	\end{tabular}
	 \caption{\label{fig:wcdm_fw0wacdm_delete1} Same as Figure \ref{fig:flcdm_delete1}, but for $w$CDM (left) and flat $w_0\,w_a$CDM (right) models. We note that the resampling techniques do not agree well with Bayesian estimates. For some parameters, large bias pushes the Jackknife contours farther from the rest. We also observe multimodal behaviour in Bootstrap estimates of some parameters. See table \ref{T:wcdm} and \ref{T:fw0wacdm} for estimates of mean, error, bias and a discussion.
    }
\end{figure}
%%%%%%%%%%%%%%%%%%
\section{Constraints on models beyond flat $\Lambda$CDM}\label{sec:constraints}
In this section, we present the constraints on other models of dark energy beyond the simplest three-parameter flat $\Lambda$CDM model, namely, $\Lambda$CDM, flat $w$CDM, $w$CDM, and flat $w_0\,w_a$CDM models, using the PPS dataset. We apply the frequentist GLS method, the Bayesian methods, {\it viz.}, MCMC and nested sampling algorithms, and the resampling techniques, namely, Jackknife delete-1, Jackknife delete-$41$, and the Bootstrap method. For GLS, we report the best-fit and the standard deviation. For the resampling methods, we report the bias-corrected estimate and standard deviation of parameters. We also report the bias. For the Bayesian analysis, we report the marginalised mean values of parameters and their standard deviations obtained using MCMC and nested sampling algorithms.
\par 
 Although the constraints on the flat $\Lambda$CDM model using the PPS dataset are already discussed in the subsection \ref{sub:diff_data}, we also include them in this section in the table \ref{T:Flatlcdm} to maintain uniformity in the presentation of our analyses. The results of $\Lambda$CDM, flat $w$CDM, $w$CDM, flat $w_0\,w_a$CDM models are shown in tables \ref{T:lcdm}, \ref{T:fwcdm}, \ref{T:wcdm}, and \ref{T:fw0wacdm} respectively. We also present the corresponding corner plots in figures \ref{fig:flcdm_delete1}, \ref{fig:lcdm_fwcdm_delete1}, and \ref{fig:wcdm_fw0wacdm_delete1}. To avoid clutter, we have plotted only results for Jackknife delete-1. Note that, unlike the contours plotted in the frequentist method, Bayesian contours are actually derived from the posterior distribution. 
 The Bayesian contours in these figures are from \texttt{PolyChord}.
 We have verified that both Bayesian methods lead to similar constraints.  
\par 
A discussion of results for each model using different methods is provided in the caption of the corresponding tables and figures. Here, we highlight some interesting aspects. First, from 2D contours, we note that Bayesian contours are more conservative across all models, except for the Jackknife contour between $h$ and $M$. We also note that the constraints on $h$ and $M$ are largely independent of the parametrisation of dark energy. Though bias correction pushes the value of $h$ to a slightly larger value, due to a larger error bar, the Planck estimate of $h\,=\,0.674$ is within $3\sigma$ (see table \ref{T:Flatlcdm}). This reduces the Hubble tension (see, for instance, \cite{DiValentino:2021izs, CosmoVerseNetwork:2025alb} and references therein). Second, from table \ref{T:lcdm}, we note that the bias correction pushes the value of $\Omega_{\rm DE}$ to a larger value which corresponds to a spatially flat universe. Moreover, from table \ref{T:fwcdm}, we note that the Jackknife delete-1 bias-corrected value of $w_0$ is closer to $-1$ than other estimates. We also note that, unlike other models, the bias-to-error ratio is small for all parameters for flat $w$CDM model. For this model, constraints from all methods are consistent with dark energy as a cosmological constant. Finally, we find that for five parameter models, the constraints from the resampling techniques tend to depart from other methods. Moreover, we see multimodality in contours obtained with Bootstrap. 
%%%%%%%%%%%%%%%%%%%%%%%%%
\section{Summary and Discussion}\label{sec:discussion}
%%%%%%%%%%%%%%%%%%%%%%%%%
Bayesian methods are the go-to technique for parameter estimation in Cosmology. Frequentist methods have rarely been used. In this article, we applied frequentist resampling techniques, namely Jackknife and Bootstrap, together with GLS, to constrain models of dark energy using the PPS dataset and compared them with Bayesian methods, {\it viz.} MCMC and nested sampling. We studied flat $\Lambda$CDM, $\Lambda$CDM, flat $w$CDM, $w$CDM, and flat $w_0\,w_a$CDM models. We presented these models,  the PPS dataset and various statistical techniques in sections \ref{sec:background}, \ref{sec:dataset}, and \ref{sec:techniques}, respectively. 
We believe these sections provide a concise and good description of these topics and would be a valuable addition to the literature. The analysis is described in sections \ref{sec:calibration} and \ref{sec:constraints}. 
\par 
In section \ref{sec:calibration}, we calibrated these methods by applying them to a near-ideal mock dataset, assuming a fiducial flat $\Lambda$CDM cosmology, and then to three versions of the PantheonPlus data. We found that Jackknife's strength is its ability to estimate bias. When the data were clean, the bias-corrected Jackknife estimates were the closest to the fiducial values of the parameters with a very small error (see table \ref{mock_covariance_corrected}). This is further corroborated by the fact that, when the degeneracy between $h$ and $M$ was broken by fixing $M$, estimates from other methods became equal to the Jackknife estimate obtained without breaking this degeneracy (see table \ref{mock_covariance_corrected_fixM}). On the other hand, when we considered PP dataset, which is relatively less clean and has more degeneracy between $h$ and $M$,  Jackknife estimated a very large bias (see table \ref{T:diffdata}).  The Jackknife estimates are largely consistent with those of other methods when either the redshift range was limited to $z>0.01$ or when the degeneracy between $h$ and $M$ was broken by including SH0ES data. 
With mock data, though the error estimated by Bootstrap is similar to that of Jackknife, it underestimates the bias. This makes its estimate far removed from fiducial values. With different versions of PantheonPlus data, its estimates are similar to those of Bayesian. 
\par 
The constraints on parameters of different models from PPS data obtained with different methods are described in section \ref{sec:constraints}. We note that the estimates of the parameters and their standard deviation obtained with GLS and Jackknife do not assume an underlying probability distribution. However, the corner plot assumes a Gaussian and $F$ distribution respectively. Also, our implementation of Bootstrap relies on a Gaussian kernal density estimator. Whereas Bayesian constraints and the contours are derived from the posterior probability distribution. 
Tables \ref{T:Flatlcdm}, \ref{T:lcdm}, \ref{T:fwcdm}, \ref{T:wcdm}, and \ref{T:fw0wacdm} lists the different constraints on flat $\Lambda$CDM, $\Lambda$CDM, flat $w$CDM, $w$CDM, and flat $w_0\,w_a$CDM respectively. We have given the corner plots of these three, four and five parameter models in figures \ref{fig:flcdm_delete1}, \ref{fig:lcdm_fwcdm_delete1}, and \ref{fig:wcdm_fw0wacdm_delete1}, respectively. For a detailed analysis of these results, we refer to the captions of these tables and figures. Here, we discuss a few interesting observations. Estimates of $h$ and $M$ obtained from PPS are largely independent of the cosmological model. Jackknife estimates a larger error in $h$ and $M$. We also observe a large correlation between these parameters with the Jackknife. Large error in $h$ makes the Planck estimate within $3\sigma$. We note that this reduces the Hubble tension. 
When we work with PP[$z>0.01$], Bayesian methods estimate $h\,=\,0.729\,\pm\,0.030$ which makes the Planck estimate within $2\sigma$. 
Surprisingly, with the same data, bias-corrected Jackknife estimates $h\,=\,0.678\, \pm\, 0.052$ which completely resolves the Hubble tension (see table \ref{T:jackknife_comp}). 
Another interesting result with Jackknife is the effect on parameter estimates due to bias correction. We note that for $\Lambda$CDM model, bias correction makes estimate of $\Omega_{\rm DE}$ larger, implying a more flat universe (see table \ref{T:lcdm}). Further, for flat $w$CDM, bias correction pushes the value of $w_0$ closer to $-1$ (see table \ref{T:fwcdm}). 
Finally, it seems that PPS data is insufficient to constrain five-parameter models. This is reflected in large error bars for all methods and large bias in resampling techniques. We also observe multimodality in 1D probability distributions and the confidence contours obtained with Bootstrap. 
\par 
With near-ideal mock data, the estimates from all methods agreed well only when the degeneracy between $h$ and $M$ was broken. So, we believe disagreement between methods could indicate insufficient data or unbroken degeneracies. Hence, it is advisable to use multiple methods to test the correctness of parameter constraints. This is the lesson that we learn from the analysis described in this article. We conclude this article with a discussion of a few aspects that need further investigation. First, the Bootstrap fails to recover the fiducial values of the parameters from a near-ideal mock data set, while its estimates are comparable to those of others using PPS data. 
Second, most interesting observation with Jackknife was that its bias corrected estimates reduces Hubble tension. We showed that when we use Jackknife, the Planck estimate is within $3\sigma$ of that obtained with PPS data and more surprisingly the Hubble tension is completely resolved when we work with PP data restricted to $z > 0.01$. It is important to investigate this as well as other constraints further, perhaps with more and better data \cite{Rubin:2023jdq, DES:2025sig, Rigault:2024kzb}. Finally, in this article, we have demonstrated how to use resampling techniques with supernovae data, it would be interesting to extend this to other cosmological data sets. 
%%%%%%%%%%%%%%%%%%%%%%%%%%
% \bibliographystyle{unsrt}
 \bibliography{Refs}
%%%%%%%%%%%%%%%%%%%%%%%%%%
\end{document}